\documentclass{mem}
\usepackage{natbib}\usepackage{txfonts}\usepackage{balance}
\usepackage{graphicx}
\usepackage[a4paper]{hyperref}
\idline{0}{0}
\begin{document}
\def\teff{$T\rm_{eff }$}
\def\kms{$\mathrm {km s}^{-1}$}

\title{
$\beta$~~Cephei stars in the inner part of the Galaxy
}

   \subtitle{}

\author{
A.\,Narwid, 
Z.\,Ko{\l}aczkowski,
A.\,Pigulski,
\and 
T.\,Ramza
}

\offprints{A.\,Narwid}
 
\institute{
Instytut Astronomiczny Uniwersytetu Wroc{\l}awskiego,
Kopernika 11, 51-622 Wroc{\l}aw, Poland,
\email{narwid@astro.uni.wroc.pl}
}

\authorrunning{Narwid et al.}

\titlerunning{Galactic $\beta$~~Cephei stars}

\abstract{From the Fourier analysis of the catalogue of $\sim$200,000 variable candidates in the OGLE-II
Galactic fields, we found a sample of about 230 short-period low-amplitude variable stars.  From their
position in the colour-magnitude diagram and the observed periods (multiple in most cases), we identify
the stars as a mixture of $\beta$~Cephei and $\delta$~Scuti stars.  $\beta$~Cephei stars from this
sample are located in the Galactic disk at distances from 3 to 6~kpc.  Many of them shows large range of the
excited periods, an indication of high metallicity.  We estimate that even a half of the sample 
of 230 short-period variables we found, can be $\beta$~Cephei stars.  The periods alone, however, are 
rarely sufficient to distinguish between both types of pulsators.  We point out, how this can be done observationally.

\keywords{Stars: $\delta$~Scuti, Stars: pulsations}
}
\maketitle{}

\section{Introduction}
The analysis of the OGLE-II observations by means of the image subtraction method resulted in a 
catalogue of over 200,000 variable star candidates published by \citet{wozn02}.  
The observations were carried out in the years 1997--2000 in 49 Galactic fields covering $\sim$11 
square degrees in the sky.  In the search for short-period pulsators in the Galaxy, we analyzed 
the data from this catalogue.  The analysis yielded a lot of candidates for $\beta$~Cephei, $\delta$~Scuti, 
SPB and $\gamma$~Doradus stars.  In this paper, we present global properties of a sample of $\sim$230 
stars with shortest periods (exluding high-amplitude $\delta$~Scuti stars, see Pigulski et al., these proceedings).   
This sample consists 
mainly of a mixture of $\beta$~Cephei and low-amplitude $\delta$~Scuti stars.  

\section{Analysis}
The analysis consisted of an automatic extraction of up to five periodic terms for all 
stars in the catalogue with consecutive prewhitening followed by an automatic classification 
based upon the periods, amplitudes, and Fourier coefficients (for stars with detected 
harmonics or subharmonic).   Then, for stars selected in this way, a detailed analysis 
was performed in an interactive way.  As a result, we selected about 230 short-period 
stars that are good candidates for $\beta$~Cephei and low-amplitude $\delta$~Scuti stars.  
For these stars, we searched for the photometry from the MACHO survey.  It turned out 
that 94 stars from this sample have the MACHO photometry available.  However, it is of 
low quality for some stars, so that we combined OGLE-II and MACHO photometry for only 67 
stars in our sample.  The analysis of the combined photometry results in a better 
resolution, lower detection threshold and practically removes the ambiguity in the 
frequencies of the extracted terms.

\begin{figure}[t!]
\resizebox{\hsize}{!}{\includegraphics[clip=true]{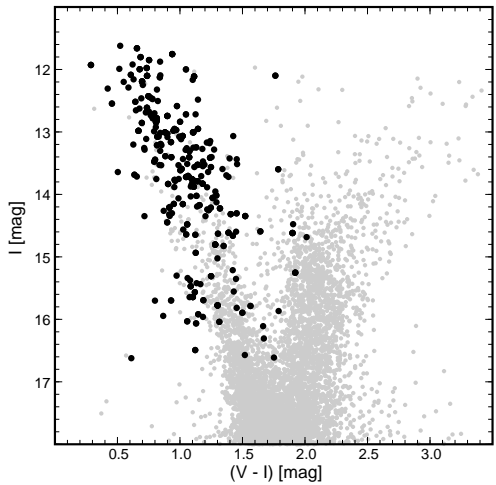}}
\caption{
\footnotesize
Colour-magnitude diagram for over 5000 stars from the OGLE-II SC11 field (gray dots) 
and a sample of $\sim$230 short-period stars (filled circles) found in the OGLE-II catalogue of \citet{wozn02}.   
The latter are mainly $\beta$~Cephei and low-amplitude $\delta$~Scuti stars.}
\label{cmd}
\end{figure}

\begin{figure}[t!]
\resizebox{\hsize}{!}{\includegraphics[clip=true]{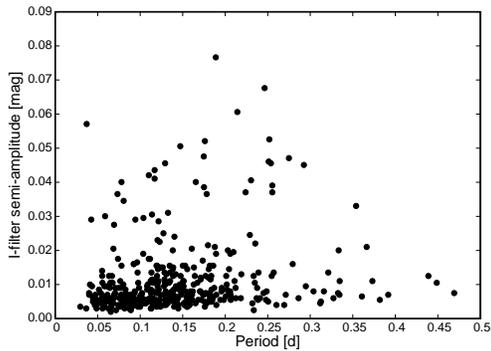}}
\caption{
\footnotesize
$I$-filter semi-amplitudes plotted against the period for all modes detected in the sample of 230 stars shown in Fig.\ref{cmd}. }
\label{perampl}
\end{figure}

\section{Results}
The main results of our study can be summarized as follows:
\begin{enumerate}
\item In the OGLE-II catalogue of variable star candidates \citep{wozn02} we found about 
230 low-amplitude stars with periods shorter than 0.5~d that are good candidates 
for $\beta$~Cephei and low-amplitude $\delta$~Scuti stars.  
In the colour-magnitude diagram (Fig.~\ref{cmd}), they populate the branch of main-sequence stars.  
Their $I$ magnitudes range from the OGLE-II saturation limit of $I \sim$ 11.5 down to magnitude 16.5. 
\item The periods of the bulk of stars in our sample range between 0.04 and 0.2~d (Fig.~\ref{perampl}).  
Since the periods of modes excited in both $\beta$~Cephei and $\delta$~Scuti stars cover this range, 
the two types of pulsators cannot be distinguished merely from their periods.  
We are going to carry out the $UBV$ photometry or/and low-resolution spectroscopy of stars 
from our sample. This will allow us to classify definitely these stars.
\item Almost all other variable stars with periods shorter than 0.5~d were excluded in the process 
of automatic classification.  However, a small-amplitude RR Lyrae and W~UMa stars may contaminate our sample, especially for
periods longer than $\sim$0.25 d.  We estimate, however, that this contribution does not exceed 10\%.
\item Half of the stars from our sample (114) shows more than one period of variability.  
Up to eight modes in a single star were found in the combined OGLE-II and MACHO data.  
\item In some stars, equidistant in frequency tri\-plets, an indication of rotational splitting, were found. 
For many stars, the range of periods of the excited modes is very wide.  This may be an indication of 
the larger-than-average metallicity expected in young population of stars in the inner regions of our Galaxy.
\item All the information given above shows clear\-ly that our sample includes many $\beta$~Cephei stars located 3--6 kpc away.
\end{enumerate}

\begin{acknowledgements}
This work has been supported by the KBN grant No.~1\,P03D\,016\,27.
\end{acknowledgements}

\bibliographystyle{aa}

\end{document}